\documentclass[twocolumn,showpacs,preprintnumbers,amsmath,amssymb]{revtex4}

\usepackage{amsmath}
\usepackage{amsfonts}
\usepackage{amstext}     
\usepackage{amssymb}
\usepackage{bbm}
\usepackage{bm}
\usepackage{mathrsfs}
\usepackage{color}
\usepackage{graphicx}
\usepackage{dcolumn}
\usepackage{bm}

\begin{document}

\title{Nonequilibrium Quantum Phase Transitions in the Ising Model}
\author{V. M. Bastidas}
\email{victor@physik.tu-berlin.de}

\author{C. Emary}
\author{G. Schaller}
\author{T. Brandes}

\affiliation{%
Institut f\"ur Theoretische Physik, Technische Universit\"at Berlin, Hardenbergstr. 36, 10623 Berlin, Germany}%

\begin{abstract}
We establish a set of nonequilibrium quantum phase transitions in the Ising model driven under monochromatic nonadiabatic modulation of the transverse field.
We show that besides the Ising-like critical behavior, the system exhibits an anisotropic transition which is absent in equilibrium. The nonequilibrium quantum phases correspond to states which are synchronized with the external control in the long-time dynamics. 
\end{abstract}


\maketitle
One of the more intriguing hallmarks of many-body systems is that at zero temperature quantum fluctuations can drive the system to a drastic change of state, commonly known as a quantum phase transition (QPT). A paradigmatic model for QPTs is the one-dimensional Ising model \cite{Sachdev}. Recently, experimental realizations of  one-dimensional spin chains have been suggested, where a quantum simulation of the system close to the phase transition is possible, and a wide freedom on the control of the parameters is achieved \cite{JohnsonRose,Barreiro, Simon, Coldea,Mostame,Trotzky}.
The quantum control of many-body systems by a driving field has attracted considerable interest, both theoretical and experimental, with workers from very different communities beginning to look at driven models \cite{Altland,Polkovnikov,Zurek,Hangui,Goren1,Holthaus,Goren2,Arimondo,Lindner,Tanaka}.
The possibility of manipulating the quantum state of a system by means of a classical external control allows  one to explore novel states of matter and effective interactions which are absent in equilibrium \cite{Lindner,Tanaka, Bastidas3, Zoller}. 
In the presence of an external control, quantum resonances and symmetries play an important role \cite{Shirley, Nori, dittrich,Grifoni}. In particular, as a consequence of a generalized parity in the extended Hilbert space \cite{Shirley}, under the effect of periodic driving the tunneling can be slowed down or totally suppressed 
in a perfect coherent way, a phenomenon commonly referred to as coherent destruction of tunneling (CDT) \cite{CDT1,CDT2}. Rather recently, the extension of this concept to many-body systems has been addressed
in the context of the Mott-insulator-superfluid transition in ultracold systems both theoretically \cite{Holthaus} as well as experimentally \cite{Arimondo}, and in a two-mode Bose-Hubbard model with time-dependent self-interaction strength \cite{Hangui}.

The dynamics of one-dimensional spin chains has been addressed extensively when the system is driven slowly through the critical point \cite{Dziarmaga, Mukherjee1, Mukherjee2}, where there is a diverging relaxation time and correlation length, and the dynamics cannot be adiabatic in the thermodynamic limit. As a consequence of this, the final state of the system consists of ordered domains whose finite size depend upon the velocity of the parameter variation \cite{KibbleZurek}. A nontrivial oscillation of the magnetization \cite{Raedt} and the connection between symmetry and CDT \cite{Miyashita} has been investigated in a finite size periodically-driven Ising model. Furthermore, under the effect of a nonadiabatic external control of the transverse field, the Ising chain exhibits dynamical freezing of the response \cite{Das, Dasgupta}, and synchronization with the external driving in the asymptotic dynamics as a consequence of destructive interference in time \cite{Santoro}.

Our aim in this paper is to describe the nonequilibrium behavior of a one-dimensional Ising model under the effect of a nonadiabatic monochromatic transverse field from the perspective of quantum criticality.
In particular, we describe the dynamics by means of an effective Hamiltonian which simulates the dynamics of an undriven system. We show that in the asymptotic dynamics the nonequilibrium quantum phases correspond to states of the system which are synchronized with the driving. In contrast to previous works \cite{Das, Dasgupta, Santoro}, however, we describe the role of many-body CDT in the critical behavior by investigating signatures of criticality both in the laboratory frame as well as in the rotating frame.

The paper is organized as follows: In Sec. \ref{SectionI} we discuss the equations of motion and describe quantum resonances by considering the excitation spectrum of the undriven system. In Sec. \ref{SectionII} we describe the physics of the system by means of the rotating wave approximation (RWA) and discuss signatures of criticality based on the description of the quasienergy dispersion and CDT. In Sec. \ref{SectionIII} we describe the quantum dynamics in the laboratory frame by considering the quantum evolution of the system when it is initialized in a paramagnetic ground state. Furthermore, we study signatures of criticality by considering cycle-averaged expectation values of physical observables. Finally, a discussion of the results is presented in Sec. \ref{SectionIV}.

\section{Quantum resonances in the Ising model. \label{SectionI}}
The periodically driven transverse Ising model describes the dynamics of $N$ interacting two-level systems in a time-dependent transverse local field \cite{Das, Dasgupta, Santoro}
\begin{eqnarray}
      \label{drivenIsing}
            \hat{H}(t)&=& -g(t)\sum_{i=1}^{N}\sigma_i^x-J\sum_{i=1}^N \sigma_i^z \sigma_{i+1}^z
      ,
\end{eqnarray}
where $\sigma_{i}^{\alpha}$ are Pauli matrices at the $i$th site and we assume periodic boundary conditions $\sigma_{1}^{\alpha}=\sigma_{N+1}^{\alpha}$ for $\alpha \in \{x,y,z\}$.
In the following we shall consider $J>0$, and a monochromatic modulation of the transverse field with a static contribution 
$ g(t)= g_{0} + g_{1} \cos \Omega t$. Associated with the Hamiltonian equation \eqref{drivenIsing} is a conserved parity $\hat{\Pi}$, such that $[\hat{H}(t),\hat{\Pi}]=0$, which is given by
\begin{equation}
      \label{IsingParity}
            \hat{\Pi}=\bigotimes^{N}_{i=1}\sigma^{x}_{i}.
\end{equation}
In the thermodynamic limit $N\rightarrow \infty$, the undriven Ising model $(g_{1}=0)$ exhibits a second-order QPT at $g_{0}^c=J$ from a symmetric paramagnetic phase ($g_{0}>J$)  to a symmetry-broken ferromagnetic phase ($g_{0}<J$) \cite{Sachdev}. 
Our aim in this paper is to study the new aspects of criticality under the effect of driving. \\

In this section we provide the basics on the formalism used to describe the time-dependent Ising model. In particular, we find a resonance condition related to $m$-photon processes under the effect of driving.
%
\subsection{The dynamic Bogoliubov-de Gennes equations}
In this paper we consider the restriction of the Hamiltonian equation \eqref{drivenIsing} to the subspace with even $(+)$ number of fermionic quasiparticles (see Appendix \ref{AppendixA}).
After a Jordan-Wigner transformation and a discrete Fourier transform of Hamiltonian equation \eqref{drivenIsing} we obtain 
\begin{eqnarray}
      \label{FourierIsing}
             \hat{H}(t) &=& \sum_{k>0}\left\{2\left[g(t)-J\cos k\right](c_k^{\dagger}c_k+c_{-k}^{\dagger}c_{-k})-2g(t)\right\}  
             \nonumber \\&&
             +\sum_{k>0}2 J\sin k(c_k^{\dagger}c_{-k}^{\dagger}+c_{-k} c_{k})
              = \sum_{k>0} \hat{H}_{k}(t)
      ,
\end{eqnarray}
where $c_k^{\dagger}$ and $c_k$ are fermionic operators \cite{Sachdev}. For finite size $N$ of the spin chain, the quasimomentum is restricted to $k \in \{\pm\frac{\pi}{N},\pm\frac{3\pi}{N},\ldots,\pm \frac{(N-1)\pi}{N}\}$. In the following, we focus on the thermodynamic limit $N\rightarrow\infty$, where we have $-\pi\leq k \leq \pi$. 

Even if we prepare the system initially in a ground state of the undriven model, under the effect of nonadiabatic external driving, the system will experience transitions to excited states. Rather recently, a formalism has been developed to deal with this kind of dynamical situation \cite{Goren2,Dziarmaga,Mukherjee1,Mukherjee2}.
The idea is based on the fact that Hamiltonian equation \eqref{FourierIsing} conserves momentum and parity. As a consequence, we can use the BCS ansatz for the evolution of the quantum state of the system
\begin{equation}
      \label{BCSAnsatz}
            |\psi,t\rangle=\bigotimes_{k>0}\left[u_{k}(t)|1_{-k},1_{k}\rangle+v_{k}(t)|0_{-k},0_{k}\rangle\right]
            ,
\end{equation}
which implies that for a given quasimomentum $k$, the quantum evolution is restricted to the Nambu subspace $\{|1_{-k},1_{k}\rangle,|0_{-k},0_{k}\rangle\}$, consisting of doubly occupied $|1_{-k},1_{k}\rangle$ and  unoccupied $|0_{-k},0_{k}\rangle$ states of $\pm k$ fermions \cite{Mukherjee1,Mukherjee2,Chakravarty}.\\
The matrix representation of the operator $\hat{H}_{k}(t)$ in the Nambu subspace is given by
\begin{equation}
      \label{TimeDepBogoliubovHamiltonian}
H_{k}(t)=\left(%
\begin{array}{cc}
\mu(t)-2\omega_k & \Delta_k\\
 \Delta_k &  -\mu(t) 
\end{array}
\right),
\end{equation}
where $H_{k}(t)$ is the Bogoliubov-de Gennes (BdG) Hamiltonian,
$\omega_k = 2J\cos k$, $\Delta_k = 2J\sin k$, and $\mu(t) = 2g(t)$.
By defining the spinor $\Psi_{k}^{\dagger}(t)=(u^{*}_{k}(t),v^{*}_{k}(t))$, and considering the BCS ansatz Eq \eqref{BCSAnsatz}, it is possible to show that the coefficients $u_{k}(t)$ and $v_{k}(t)$ should satisfy the differential equation
\begin{equation}
      \label{DynBogoliubovdeGennes}
            i\frac{d}{dt}\Psi_{k}(t) = H_{k}(t)\Psi_{k}(t)
      ,
\end{equation}
which constitutes the dynamical version of the Bogoliubov-de Gennes equation \cite{Dziarmaga}. At this point we have translated the many-body problem into the solution of the time-dependent Schr\"{o}dinger equation for an effective two-level system.\\ 
Under periodic driving, the Floquet theorem states that the solution of equation \eqref{DynBogoliubovdeGennes} can be written
as
\begin{eqnarray}
      \label{ExpansionFloquet}
      \Psi_{k}(t) = A_{+} e^{-i\varepsilon_{k}^{(+)} t}\Phi_{k}^{(+)}(t)+A_{-} e^{-i\varepsilon_{k}^{(-)} t}\Phi_{k}^{(-)}(t)
      ,
\end{eqnarray}
where $\Phi_{k}^{(\pm)}(t)$ denote the Floquet modes corresponding to the quasienergies $-\Omega/2\leq\varepsilon_{k}^{(\pm)}\leq\Omega/2$.
Furthermore, in the extended Hilbert space $\mathcal{R}\otimes\mathcal{T}$, where $\mathcal{R}$ is the Hilbert space of square integrable functions and $\mathcal{T}$ is the space of time-periodic functions,
the Floquet states satisfy the eigenvalue problem
\begin{equation}
      \label{FloquetBogoliubovdeGennes}
            \mathcal{H}_{k}\Phi^{(\lambda)}_{k}(t)=\varepsilon^{(\lambda)}_{k} \Phi^{(\lambda)}_{k}(t) 
      ,
\end{equation}
where $\lambda \in \{+,-\}$, $\mathcal{H}_{k} = H_{k}(t)-i\hat{\mathbbm{1}}_{k}\frac{\partial}{\partial t}$ is the Floquet- Bogoliubov-de Gennes (FBdG) Hamiltonian, $\varepsilon^{(\lambda)}_{k}$ are the quasienergies, and the Floquet modes $\Phi_{k}^{(\lambda)}(t+T)=\Phi_{k}^{(\lambda)}(t)$  have the same period $T=2\pi/\Omega$ as the external driving \cite{Shirley, Nori, dittrich,Grifoni}.

\subsection{Resonance conditions}
\begin{figure}
\centering
\includegraphics[width=0.81\linewidth]{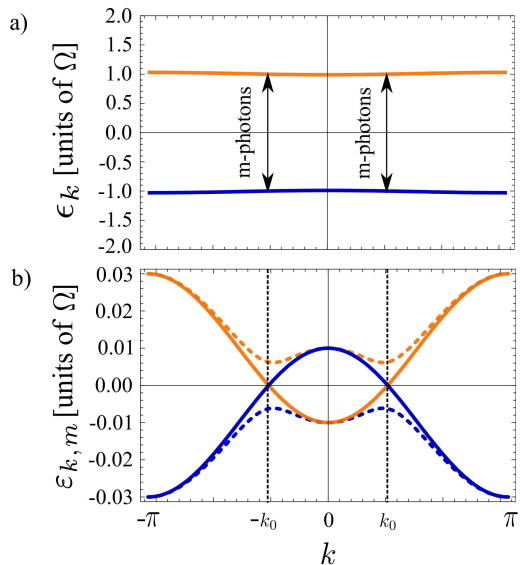}
\caption{(Color online)
  \label{Fig1}
 (a) Typical gapped energy spectrum $\pm \epsilon_k$ of the undriven system corresponding to the paramagnetic phase $g_0 \gg J$. In this case, the energy dispersion is slightly curved because the curvature is proportional to the spin-spin interaction strength $J$. (b) The continuous lines depict the spectrum $\pm \epsilon_k$ when it is folded  into the first Brillouin zone, the crossing at $k=\pm k_0$ is related to a two-photon resonance. The dashed lines represent the quasienergy spectrum $\pm\varepsilon_{k,m}$ for a driving amplitude $g_1/\Omega=1$. The parameters are $m=2, J/\Omega=0.01$, and $g_{0}/\Omega=0.505$.
}
\end{figure}

In the thermodynamic limit, the Ising model is characterized by an infinite set of collective excitations. Under the effect of an external driving, the possibility of multiphotonic resonances arises \cite{Shirley,dittrich,Grifoni}. To study such quantum resonances, let us consider the system in the absence of driving $g_{1}=0$.
In this case, the Floquet modes and the quasienergies become the stationary states $\phi_{k}^{\pm}$ and the excitation spectrum $\epsilon_{k}^{(\pm)}=-\omega_k\pm \epsilon_{k}$ of the undriven system, respectively \cite{dittrich,Grifoni}. Therefore, the solution of equation \eqref{DynBogoliubovdeGennes} can be written in the form of equation \eqref{ExpansionFloquet} as follows:
\begin{eqnarray}
      \psi_{k}(t) = a_{+} e^{-i\epsilon_{k}^{(+)} t}\phi_{k}^{(+)}+a_{-} e^{-i\epsilon_{k}^{(-)} t}\phi_{k}^{(-)}
      .
\end{eqnarray}
The energy gap is given by $\Delta E_k=\epsilon_{k}^{(+)}-\epsilon_{k}^{(-)}=2\epsilon_{k}$,
where
\begin{equation}
      \label{EnergyExcitation}
            \epsilon_{k}=2\sqrt{\left(g_{0}-J\cos k\right)^2+\left(J\sin k\right)^2}
      .
\end{equation}
In the semiclassical theory of light-matter interaction, we can interpret a Floquet state as a light-matter quantum state containing a definite, though very large, number of photons \cite{Shirley}.
Multiple transitions between quantum states of the spin chain that are not directly coupled by the interaction can occur by means of intermediate states with a different number of photons present \cite{Shirley, Nori,dittrich}. In particular, $m$-photon transitions occur when the condition
\begin{equation}
      \label{EnergyGapResonanceCond}
            \Delta E_k=m\Omega     
      ,
\end{equation}
with integer $m$ is satisfied. For a parametric oscillator with fundamental frequency $\epsilon_{k}$, equation \eqref{EnergyGapResonanceCond} is the usual resonance condition \cite{Arnold}.
In Floquet theory, equation \eqref{EnergyGapResonanceCond} implies the existence of a crossing between the single-particle energy levels $\epsilon_{k}$ and $-\epsilon_{k}$ when the energy spectrum is folded into the Brillouin zone \cite{Shirley}. Such a crossing occurs at the wave vector
\begin{equation}
      \label{QuasimomentumCrossing}
            k_0=\pm\arccos\left(\frac{g^{2}_{0}+J^2-(\frac{m\Omega}{4})^2}{2g_{0}J}\right)
      ,
\end{equation}
where the resonance condition is fulfilled, as depicted in figure~\ref{Fig1} (a).  figure~\ref{Fig1} (a)
depicts the energy dispersion relation of the undriven system, and the continuous lines in figure~\ref{Fig1} (b), the corresponding folding of the energy spectrum into the first Brillouin zone $-\Omega/2\leq\epsilon_k\leq\Omega/2$. In this paper we focus on the weak spin-spin coupling limit $g_{0},\Omega \gg J$. In this limit the multiphoton resonance condition reads 
\begin{equation}
      \label{ResWeakCondition}
            g_{0}=\frac{m\Omega}{4}
      .
\end{equation}

Such resonance condition will be used in the next section to perform a description of the system based on an effective time-independent Hamiltonian which is valid for parameters close to a multiphotonic resonance.
\section{Physics in the rotating frame \label{SectionII}}
The dynamical BdG equations allow us to investigate the full quantum evolution of the driven system. As we are interested in the asymptotic quantum dynamics and the description of critical signatures, it is convenient to describe the dynamics of the system in a rotating frame. In the weak spin-spin interaction limit, it is possible to neglect the fast oscillations in the rotating frame, and a description of the system based on the description for time-independent systems is possible via an approximate effective Hamiltonian.
\subsection{The rotating wave approximation and the effective Hamiltonian}
Motivated by the $m$-photon resonance condition in the weak spin-spin coupling limit, Eq. \eqref{ResWeakCondition}, we perform a description of the system based on the rotating wave approximation \cite{Nori}. Let us perform a unitary transformation of Hamiltonian equation \eqref{drivenIsing} into a convenient rotating frame via the unitary operator
\begin{align}
      \label{UnitaryIsingRot}
            \hat{U}^{(m)}(t) &=
            \exp \left( i\alpha_{m}(t)\sum_{i=1}^{N}\sigma_i^x\right)=\prod_{k>0} \hat{U}_k^{(m)}(t)
            \nonumber \\
            =& \prod_{k>0} \exp \left[-2i\alpha_{m}(t)(c_k^{\dagger}c_k+c_{-k}^{\dagger}c_{-k}-1)\right]
      ,
\end{align}
where $\alpha_{m}(t)=m(\Omega/4) t+\frac{g_{1}}{\Omega}\sin\Omega t$. 
In the rotating frame the dynamics is governed by the Hamiltonian
$\hat{H}^{(m)}(t)=[\hat{U}^{(m)}(t)]^\dagger \hat{\mathcal{H}} \hat{U}^{(m)}(t)$, where $\hat{\mathcal{H}}=\hat{H}(t)-i \frac{\partial}{\partial t}=\sum_{k>0}\hat{\mathcal{H}}_k=\sum_{k>0}\left[\hat{H}_k(t)-i\hat{\mathbbm{1}}_{k} \frac{\partial}{\partial t}\right]$ is the Floquet Hamiltonian. The explicit form of this operator is given by
\begin{widetext}
\begin{eqnarray}
      \label{HamiltonianRotFrameSI}
	    \hat{H}^{(m)}(t)&=& -\delta^{(m)}\sum_{i=1}^{N}\sigma_i^x-\frac{J}{2}\left\{1+\cos[4\alpha_{m}(t)]\right\}\sum_{i=1}^N \sigma_i^z \sigma_{i+1}^z	   
	    -\frac{J}{2}\left\{1-\cos[4\alpha_{m}(t)]\right\}\sum_{i=1}^N \sigma_i^y \sigma_{i+1}^y
	    \nonumber \\&&
            +\frac{J}{2}\sin[4\alpha_{m}(t)]\sum_{i=1}^N \sigma_i^z \sigma_{i+1}^y            
            +\frac{J}{2}\sin[4\alpha_{m}(t)]\sum_{i=1}^N \sigma_i^y \sigma_{i+1}^z
     ,
\end{eqnarray}
\end{widetext}
where the detuning $\delta^{(m)}=g_{0}-m (\Omega/4)$ describes how far the system is from resonance, and $m$ is an integer that denotes the order of the multiphotonic resonance \cite{Shirley}. By using the identity
\begin{equation}
       \label{ExpBessel} 
	     \exp(iz\sin\Omega t)=\sum_{l=-\infty}^{\infty}\mathcal{J}_{l}(z)\exp(il\Omega t) 
       ,
\end{equation}
where $\mathcal{J}_{l}(z)$ is the  $l$th-order Bessel function \cite{Abramowitz},
the Hamiltonian equation \eqref{HamiltonianRotFrameSI} can be written in the form
\begin{equation}
      \label{IsingHamiltonianFourier}
             \hat{H}^{(m)}(t)=\sum_{n=-\infty}^{\infty} \hat{h}^{(m)}_{n}\exp{(i n \Omega t)}     
   .
\end{equation}
In analogy with the standard RWA of quantum optics, we obtain an approximate Hamiltonian to describe the $m$th resonance by neglecting all the terms in $\hat{H}^{(m)}(t)$ with oscillatory time dependence:  $\hat{H}^{(m)}(t) \approx  \hat{h}^{(m)}_{0} $. This approximation is valid as long as the condition
\begin{equation}
      \label{ValidityRWA}
            \delta^{(m)},J\mathcal{J}_{m}\left(\frac{4g_{1}}{\Omega}\right) \ll\Omega
\end{equation}
holds \cite{Nori}.\\
Finally, we obtain the time-independent effective Hamiltonian 
\begin{equation}
      \label{mResonaceIsingXY}
             \hat{h}^{(m)}_{0} = -\delta^{(m)}\sum_{i=1}^{N} \sigma_i^x-\sum_{i=1}^N (J^{(m)}_z \sigma_i^z\sigma_{i+1}^z+J^{(m)}_y\sigma_i^y\sigma_{i+1}^y)
      ,
\end{equation}
where the parameters $J^{(m)}_z=\frac{J}{2}[1+(-1)^m\mathcal{J}_{m}(\frac{4 g_{1}}{\Omega})]$ and $J^{(m)}_y=\frac{J}{2}[1-(-1)^m\mathcal{J}_{m}(\frac{4 g_{1}}{\Omega})]$ denote effective anisotropies in the rotating frame. Interestingly, the effective Hamiltonian equation \eqref{mResonaceIsingXY} corresponds to an exactly solvable model, i.e., it is unitarily equivalent to the $XY$ anisotropic spin chain in a transverse field \cite{Mattis, Barouch, Bunder}. However, in our case, the anisotropies depend upon both the order $m$ of the resonance  as well as the driving amplitude $g_1$. Therefore, the driving amplitude of the local field now plays the role
of a new parameter that influences the criticality of the system. 
\subsection{Signatures of criticality in the rotating frame}
Under the RWA, the Hamiltonian $\hat{h}^{(m)}_{0}$ 
and the Floquet Hamiltonian $\hat{\mathcal{H}}=\hat{H}(t)-i \frac{\partial}{\partial t}$ are isospectral operators, which implies that the eigenvalues of the effective Hamiltonian correspond to the quasienergies. As we show in Appendix \ref{AppendixB}--- similarly to the Ising model \cite{Sachdev}--- after a Jordan-Wigner transformation, and a discrete Fourier transform, the Hamiltonian equation \eqref{mResonaceIsingXY} can be written as follows:
\begin{align}
      \label{FourierIsingEffective}
             \hat{h}^{(m)}_{0}&= \sum_{k>0}\left[(2\delta^{(m)}-\omega_k)(c_k^{\dagger}c_k+c_{-k}^{\dagger}c_{-k})-2\delta^{(m)}\right]  
             \nonumber \\
             & +\sum_{k>0}(-1)^{m}\Delta_k\mathcal{J}_{m}\left(\frac{4 g_{1}}{\Omega}\right)(c_k^{\dagger}c_{-k}^{\dagger}+c_{-k} c_{k})
             \nonumber \\
             &= \sum_{k>0} \hat{h}^{(m)}_{0,k}
      .
\end{align}
The matrix representation of $\hat{h}^{(m)}_{0,k}$ in the Nambu subspace is given by
\begin{equation}
      \label{EffectiveBogoliubovHamiltonian}
h^{(m)}_{0,k}=\left(%
\begin{array}{cc}
2\delta^{(m)}-2\omega_k & (-1)^{m}\Delta_k\mathcal{J}_{m}(\frac{4 g_{1}}{\Omega})\\
 (-1)^{m}\Delta_k\mathcal{J}_{m}(\frac{4 g_{1}}{\Omega}) & -2\delta^{(m)}
\end{array}
\right)
.
\end{equation}
The Hamiltonian equation \eqref{FourierIsingEffective} can be diagonalized via a Bogoliubov transformation 
\begin{eqnarray}
      \label{DiagonalizedXYmodel}
             \hat{h}^{(m)}_{0} =\sum_{k>0} \varepsilon_{k,m}\left(\gamma_k^{\dagger}\gamma_k-\frac{1}{2}\right)
      ,
\end{eqnarray}
where 
\begin{equation}
      \label{QuasiEnergyExcitation}
            \varepsilon_{k,m}=2\sqrt{\left(\delta^{(m)}-J\cos k\right)^2+\left[J\mathcal{J}_{m}\left(\frac{4 g_{1}}{\Omega}\right)\sin k\right]^2}.
\end{equation}
Furthermore, the quasienergies are defined (modulus $\Omega$) by  the equation
\begin{equation}
      \label{QuasienergyDispersionEff}
            \varepsilon_{k,m}^{(\pm)}=-\omega_k\pm \varepsilon_{k,m}+\frac{m\Omega}{2}
      ,
\end{equation}
as defined in equation \eqref{FloquetBogoliubovdeGennes}. 
The quasienergy gap in the fermion picture is given by $\Delta E_{k,m}=\varepsilon_{k,m}^{(+)}-\varepsilon_{k,m}^{(-)}=2\varepsilon_{k,m}$. Therefore,
when the gap closes, modulus $\Omega$---the effective Hamiltonian---exhibits a behavior which resembles the dynamics of a critical quantum system. The dashed lines in figure~\ref{Fig1} (b) depict the quasienergy dispersion relation for $g_1\neq 0$. We observe that the driving lifts the degeneracy giving rise to an anticrossing. 
Based on the well-known results for the time-independent $XY$ model that we summarize in Appendix \ref{AppendixB}, we find that the system described by the effective Hamiltonian equation \eqref{mResonaceIsingXY} exhibits an effective nonequilibrium Ising-like QPT along the critical lines
$|\delta^{(m)}| = J$, and a nonequilibrium anisotropic QPT along the lines where $\mathcal{J}_{m}\left(\frac{4 g_{1}}{\Omega}\right) = 0$, as long as the condition $|\delta^{(m)}| < J$ holds. The gapless quasienergy excitation spectrum for parameters along the critical lines is a direct consequence of coherent destruction of tunneling \cite{CDT1,CDT2}, i.e,  of the existence of a generalized parity symmetry in the extended Hilbert space $\mathcal{R}\otimes\mathcal{T}$, where $\mathcal{R}$ is the Hilbert space of square integrable functions and $\mathcal{T}$ is the space of time-periodic functions \cite{Grifoni}.

\begin{figure}
\centering
\includegraphics[width=0.81\linewidth]{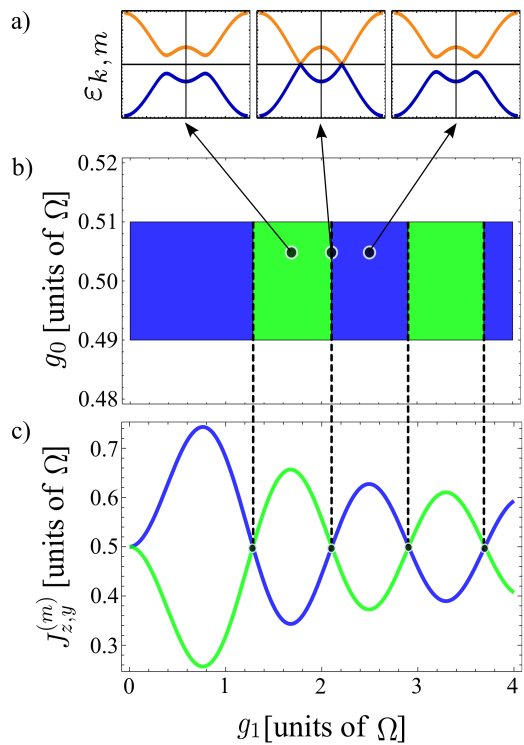}
\caption{
  \label{Fig2} (Color online)
  Nonequilibrium quantum phase transition in the driven Ising chain in a time-dependent transverse field $g(t)=g_0+g_1\cos\Omega t$. $(a)$ depicts the quasienergy dispersion relations $\pm \varepsilon_{k,m}$ for parameters in the ferromagnetic phases FM$Z$ and FM$Y$, and along a critical line.
  $(b)$ depicts the phase diagram of the nonequilibrium phase transition around  the $m=2$ resonance as a function of the driving amplitude $g_1$ and the static local field $g_0$. 
  The white zones represent the paramagnetic phase. Correspondingly, the blue (dark gray) zones represent the ferromagnetic phase FM$Z$ and the green zones (light gray) the ferromagnetic phase FM$Y$.
  $(c)$  depicts the effective asymmetries in the $z$ direction $J_{z}^{(m)}$ [blue (dark gray) curve] and $y$ direction $J_{y}^{(m)}$ [green (light gray) curve] as a function of the driving amplitude $g_1$. For this plot, we have considered $J/\Omega=0.01$.
}
\end{figure}

Figure \ref{Fig2} (a) depicts the character of the quasienergy excitation spectrum $\pm \varepsilon_{k,m}$ for parameters in the ferromagnetic phase FM$Y$, along the critical line, and in the ferromagnetic phase  FM$Z$, respectively. Figure \ref{Fig2} (b) depicts the phase diagram for the nonequilibrium QPT in the neighborhood of the two-photon resonance. The white zones in the phase diagram correspond to the effective paramagnetic phase and are defined by the inequality 
$J<|\delta^{(m)}|<|\delta_{\text{max}}^{(m)}|$, for $m=2$, where $\delta_{\text{max}}^{(m)}$ denotes the maximum detuning for which the RWA is still valid.
The anisotropic transition is characterized by two ferromagnetic phases, i.e., for $J^{(m)}_{z}> J^{(m)}_{y}$ the system is in a ferromagnetically ordered phase along the $z$ direction FM$Z$, while it is the other way around in the FM$Y$ phase. In the particular case $\delta^{(m)}=0$, the effective Hamiltonian equation \eqref{mResonaceIsingXY} is unitarily equivalent to the $XY$ model in the absence of a transverse field. Therefore, in this special case the system only exhibits the conventional anisotropic transitions
between the ferromagnetically ordered FM$Z$ and FM$Y$ phases.
In figure~\ref{Fig2} (c) we plot the effective asymmetries $J^{(m)}_{z}$ and $J^{(m)}_{y}$ as a function of the driving amplitude $g_1$ in the case of a two-photon resonance.  
\section{Physics in the laboratory frame \label{SectionIII}}
\subsection{Quantum evolution of an initial paramagnetic state}
As we previously mentioned, the eigenvalues of the effective Hamiltonian correspond to the quasienergies of the system. However, the corresponding eigenstates do not necessarily correspond to Floquet modes. In order to obtain the
Floquet modes, one should apply a unitary transformation back into the laboratory frame. In so doing, the positive-quasienergy Floquet mode around the $m$-photon resonance in the Nambu subspace is given by
\begin{align}
      \label{PositiveQuasienergyFloquet}
\Phi_{k,m}^{(+)}(t)&=e^{i\left(\frac{m\Omega}{2}t\right)}U^{(m)}_k(t)\left(%
\begin{array}{cc}
\cos(\phi_{k,m}) \\
 -\sin(\phi_{k,m}) 
\end{array}
\right)
\nonumber \\
&= \left(%
\begin{array}{cc}
e^{-i\left(2\alpha_{m}(t)-\frac{m\Omega}{2}t\right)}\cos(\phi_{k,m}) \\
 -e^{i\left(2\alpha_{m}(t)+\frac{m\Omega}{2}t\right)}\sin(\phi_{k,m}) 
\end{array}
\right),
\end{align}
and, correspondingly, the negative-quasienergy Floquet mode is given by
\begin{align}
      \label{NegativeQuasienergyFloquet}
\Phi_{k,m}^{(-)}(t) &= e^{i\left(\frac{m\Omega}{2}t\right)}U^{(m)}_k(t)\left(%
\begin{array}{cc}
\sin(\phi_{k,m}) \\
 \cos(\phi_{k,m}) 
\end{array}
\right)
\nonumber \\
&=\left(%
\begin{array}{cc}
e^{-i\left(2\alpha_{m}(t)-\frac{m\Omega}{2}t\right)}\sin(\phi_{k,m}) \\
 e^{i\left(2\alpha_{m}(t)+\frac{m\Omega}{2}t\right)}\cos(\phi_{k,m}) 
\end{array}
\right),
\end{align}
where 
\begin{equation}
      \label{BogoliuvobAngle}
            \tan(2\phi_{k,m})=\frac{-(-1)^{m}\Delta_k\mathcal{J}_{m}\left(\frac{4 g_{1}}{\Omega}\right)}{2\delta^{(m)}-\omega_k}.
\end{equation}
Now, let us investigate the quantum evolution in the laboratory frame around the $m$-photon resonance when the system is initialized in a paramagnetic state of the undriven model with all the spin polarized along the $x$ axis
\begin{equation}
      \label{InitialParamagneticState}
            |\psi_{m},0\rangle=\bigotimes_{k>0}|0_{-k},0_{k}\rangle
            .
\end{equation}
Restricted to the Nambu subspace for a given $0\leq k \leq \pi$, such an initial state corresponds to the spinor  $\Psi_{k,m}^{\dagger}(0)=(u^{*}_{k,m}(0),v^{*}_{k,m}(0))=(0,1)$, whose quantum evolution is given by
\begin{align}
      \label{EvolutionGroundState}
\Psi_{k,m}(t)&=U^{(m)}_k(t)\exp(-i h^{(m)}_{0,k} t)\Psi_{k,m}(0)
             \nonumber \\
             &=-\sin(\phi_{k,m})\Psi^{(+)}_{k,m}(t) +\cos(\phi_{k,m})\Psi^{(-)}_{k,m}(t)
      ,
\end{align}
where $\Psi^{(\pm)}_{k,m}(t)=e^{-i\varepsilon_{k,m}^{(\pm)} t}\Phi_{k,m}^{(\pm)}(t)$ denotes the Floquet states restricted to the Nambu subspace.

\subsection{The dynamics of the transverse magnetization}
By using the exact quantum evolution of the initial paramagnetic state, we are able to calculate the dynamics of physical observables in the laboratory frame.  
The transverse magnetization density $M_x(t)$ gives us information about the occurrence of a macroscopic polarization of the spins along the $x$ axis.
Let us consider the expectation value $M_x(t)=\frac{1}{N} \langle \psi_m,t| \sum_{i=1}^{N}\sigma^{x}_{i}|\psi_m,t\rangle$ close to the $m$-th resonance,
\begin{align}
      \label{EvolutionTransvMagnetization}
          M_x(t) &=-\int_{0}^{\pi}\frac{dk}{\pi}\Psi^{\dagger}_{k,m}(t)\sigma^{z}(k)\Psi_{k,m}(t)
            \nonumber \\
            &=1-2\int_{0}^{\pi}\frac{dk}{\pi}\sin^{2}(\varepsilon_{k,m}t)\sin^{2}(2\phi_{k,m})
      ,      
\end{align}
where 
\begin{equation}
       \label{SigmaZNambu}
       \sigma^{z}(k) =\frac{1}{2}\frac{\partial \mathcal{H}_{k}}{\partial g_0}=\left(%
\begin{array}{cc}
1 & 0\\
0 & -1
\end{array}
\right) ,
\end{equation}
and $\mathcal{H}_{k}$ is the FBdG Hamiltonian equation \eqref{FloquetBogoliubovdeGennes}.
\begin{figure}
\centering
\includegraphics[width=0.81\linewidth]{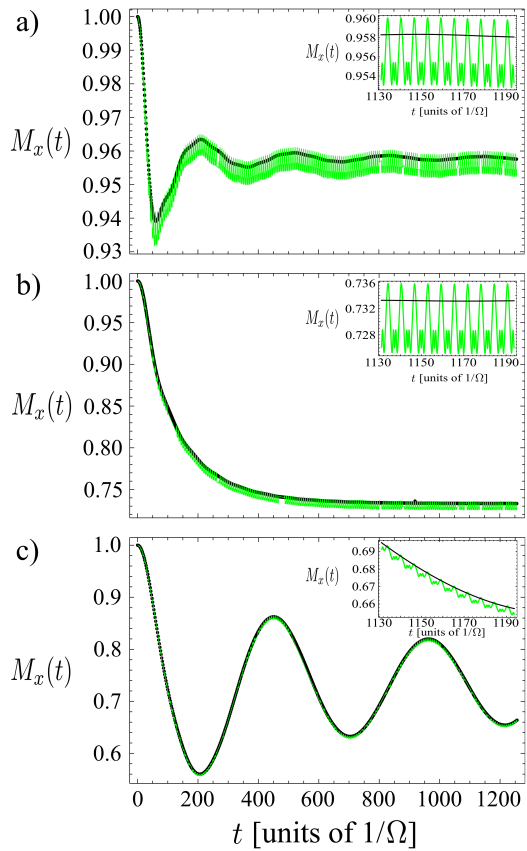}
\caption{
  \label{Fig3} (Color online)
 (a) Time evolution of the dimensionless magnetization density $M_x(t)$ in the thermodynamic limit (black curve) and exact numerical result for a finite system consisting of $N=100$ spins [green (light gray) curve] for $0<t<200T$, where $T=2\pi/\Omega$ is the period of the external driving. Time evolution for parameters corresponding to (a) the nonequilibrium paramagnetic phase $(\delta^{(m)}>J)$, for $m=2$, where $(g_1/\Omega,g_0/\Omega)=(1,0.515)$, (b) the Ising-like critical line $(\delta^{(m)}=J)$, for $m=2$, where $(g_1/\Omega,g_0/\Omega)=(1,0.510)$, and (c) time evolution for parameters corresponding to the nonequilibrium ferromagnetic phase (FM$Z$) $(\delta^{(m)}<J)$, for $m=2$, where $(g_1/\Omega,g_0/\Omega)=(1,0.505)$.
 The insets show the detail of the magnetization curves. We have considered $J/\Omega=0.01$.
}
\end{figure}
figure~\ref{Fig3} depicts the dynamics of the magnetization density in the thermodynamic limit calculated using RWA (black curve). In particular, figure~\ref{Fig3} (a) shows the dynamics for parameters in the nonequilibrium paramagnetic phase, figure~\ref{Fig3} (b) at the Ising-like critical line $\delta^{(m)}=J$, and figure~\ref{Fig3} (c) for the ferromagnetic phase FM$Z$. We observe that in the paramagnetic phase the system exhibits a stationary state which corresponds to a polarized state along the local field direction. In contrast, for parameters corresponding to the critical line and the ferromagnetic phase FM$Z$, the state is not totally polarized along this direction. Furthermore,
at the Ising-like critical line, the magnetization density does not exhibit oscillations.
The green curve (light gray) in figure~\ref{Fig3} depicts the result of exact numerical calculation (see Appendix \ref{AppendixC}) of the magnetization density for a finite size system $N=100$.\\
In this dynamical scenario the connection with criticality is not obvious. Rather, signatures of quantum criticality in the laboratory frame may appear in the asymptotic dynamics.
Let us consider now the time evolution of the expectation value of a general observable 
\begin{equation}
      \label{GeneralExpValue}
           \mathcal{O}(t)=\langle \psi_m,t|\hat{\mathcal{O}}|\psi_m,t\rangle = \int_{0}^{\pi}\frac{dk}{2\pi}\Psi^{\dagger}_{k,m}(t) \mathcal{O}_{k} \Psi_{k,m}(t).
 \end{equation}
In general, following the argument established in \cite{Santoro}, it is possible to show that $\mathcal{O}(t)=\mathcal{O}^{\text{per}}(t)+\mathcal{O}^{\text{tr}}(t)$, where
\begin{equation}
      \label{GeneralExpValuePeriodic}
          \mathcal{O}^{\text{per}}(t)=\sum_{\lambda \in \{+,-\}} \int_{0}^{\pi}\frac{dk}{2\pi} |A_{\lambda}|^2[\Phi_{k,m}^{(\lambda)}(t)]^{\dagger} \mathcal{O}_{k} \Phi_{k,m}^{(\lambda)}(t)
 \end{equation}
is the \textit{periodic} contribution to the expectation value, which corresponds to synchronization with the external driving. Here we consider $A_{+}=-\sin[\phi_{k,m}]$ and $A_{-}=\cos[\phi_{k,m}]$ . Correspondingly,
\begin{equation}
      \label{GeneralExpValueTransient}
          \mathcal{O}^{\text{tr}}(t)= 
          \int_{0}^{\pi}\frac{dk}{\pi} \text{Re}\left\{A^{*}_{+}A_{-}e^{-2i\varepsilon_{k,m}t}[\Phi_{k,m}^{(+)}(t)]^{\dagger} \mathcal{O}_{k} \Phi_{k,m}^{(-)}(t)\right\}
 \end{equation}
denotes the \textit{transient} component, which decays to zero in the long-time limit as a consequence of destructive interference in time \cite{Santoro}. Therefore, the system tends to synchronize with the external control in the long-time limit. The particular case $\delta^{(m)}=0$ for $m=0$ has been discussed in Ref. \cite{Das} in the context of freezing of the response in a many-body system. In this case, the system only exhibits the conventional anisotropic transition, which is reflected in the behavior of the
magnetization dynamics. Furthermore, the anisotropic critical lines $\mathcal{J}_0(4g_{1}/\Omega)=0$ are  related to the effect of maximal freezing discussed in Ref. \cite{Das}.
We conclude that in the asymptotic dynamics, the Floquet modes determine the quantum critical behavior, as we discuss in the next section.
\subsection{Cycle-averaged expectation values in Floquet eigenstates}
We now define cycle-averaged expectation values of physical observables.
In the case of a time-dependent Hamiltonian $\hat{H}(t)$, the energy is not conserved. Therefore, to describe signatures of the quantum phase transition in the laboratory frame we define the averaged energy $\bar{H}_{m}^{(\pm)}$ in the Floquet state $|\Psi^{(\pm)}_{m}(t)\rangle=\bigotimes_{k>0}|\Psi^{(\pm)}_{k,m}(t)\rangle$ as 
\begin{align}
      \label{GeneralAveragedEnergy}
            \bar{H}_{m}^{(\pm)}&\equiv\int_{0}^{T}\frac{dt}{T}\int_{0}^{\pi}\frac{dk}{2\pi}\left[\Psi^{(\pm)}_{k,m}(t)\right]^{\dagger}H_{k}(t)\Psi^{(\pm)}_{k,m}(t)
            \nonumber \\
            &=\int_{0}^{T}\frac{dt}{T}\int_{0}^{\pi}\frac{dk}{2\pi} \left(\varepsilon_{k,m}^{(\pm)}+\left[\Phi_{k,m}^{(\pm)}(t)\right]^{\dagger} i\frac{\partial}{\partial t}\Phi_{k,m}^{(\pm)}(t)\right)
      .      
\end{align}
\begin{figure}
\centering
\includegraphics[width=0.81\linewidth]{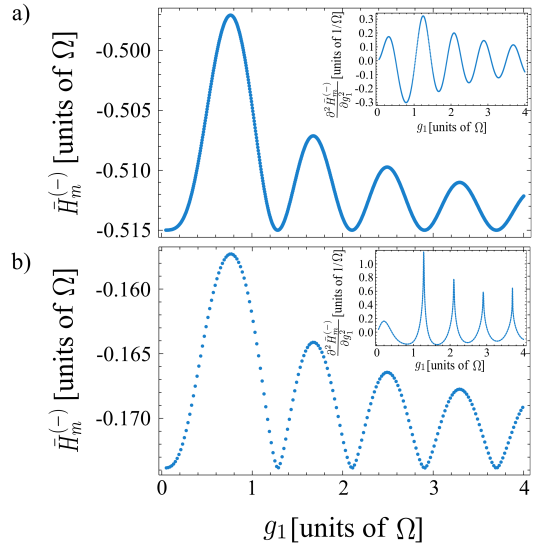}
\caption{
  \label{Fig4}(Color online)
  Cycle-averaged energy $\bar{H}_{m}^{(-)}$ as a function of the driving amplitude $g_1$.
   (a) For parameters in nonequilibrium paramagnetic phase $(\delta^{(m)}>J)$, for $m=2$, where $g_0/\Omega=0.515$. (b) For parameters inside the ladder of ferromagnetic phases $(\delta^{(m)}<J)$, for $m=2$, where $g_0/\Omega=0.505$.
   The insets depict the second derivative of the cycle-averaged energy as a function of the driving amplitude $g_1$.
   We have considered $J/\Omega=0.01$.
}
\end{figure}
By using the analytic expressions for the Floquet modes, Eqs. \eqref{PositiveQuasienergyFloquet} and \eqref{NegativeQuasienergyFloquet}, we obtain the expression
\begin{equation}
      \label{AveragedEnergies} 
	    \bar{H}_{m}^{(\pm)}= \pm \int_{0}^{\pi}\frac{dk}{2\pi}\left(\varepsilon_{k,m} + \frac{m\Omega}{2} \cos(2\phi_{k,m})\right)
      .
\end{equation}
On exact resonance $\delta^{(m)}=0$, we obtain an analytical expression for the cycle averaged energy
\begin{equation}
      \label{ExactDynamicalPhase}
            \bar{H}_{m}^{(\pm)}=\pm \frac{2 J}{\pi} E\left\{1-\left[\mathcal{J}_{m}\left(\frac{4 g_{1}}{\Omega}\right)\right]^{2}\right\}
      ,
\end{equation}
where $E[z]$ is the complete elliptic integral of the second kind (see Appendix \ref{AppendixB}). This result confirms our prediction based on the description of the system in the rotating frame (see figure~\ref{Fig2}). The cycle-averaged energy exhibits singularities at the zeros of the Bessel function, i.e., 
$\mathcal{J}_{m}\left(\frac{4 g_{1}}{\Omega}\right)=0$. This is a clear signature of a critical nonequilibrium behavior. 
\begin{figure}
\centering
\includegraphics[width=0.81\linewidth]{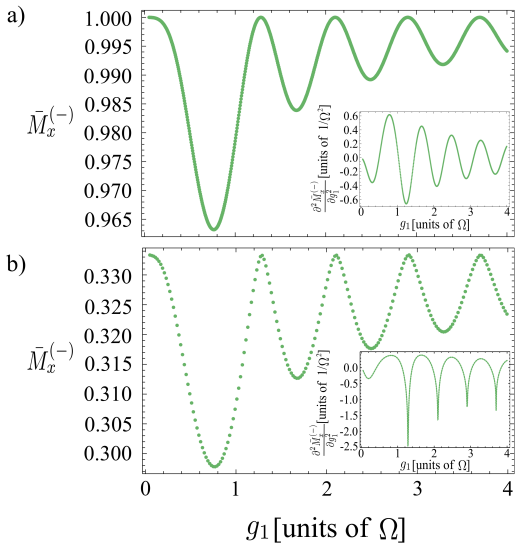}
\caption{
  \label{Fig5} (Color online)
 Cycle-averaged dimensionless magnetization $\bar{M}_{z}^{(-)}$ as a function of the driving amplitude $g_1$. 
 (a) For parameters in the nonequilibrium paramagnetic phase $(\delta^{(m)}>J)$, for $m=2$, where $g_0/\Omega=0.515$. (b) For parameters inside the ladder of ferromagnetic phases $(\delta^{(m)}<J)$, for $m=2$, where $g_0/\Omega=0.505$.
 The insets depict the second derivative of the cycle-averaged magnetization as a function of the driving amplitude $g_1$.
 The parameters are $m=2$ and $J/\Omega=0.01$.
}
\end{figure}
Finally, we calculate the cycle-averaged transverse magnetization in the Floquet mode $|\Psi^{(\pm)}_{m}(t)\rangle$. By considering the extension of the Hellmann-Feynman theorem for Floquet theory \cite{Grifoni,dittrich}, we can compute the cycle-averaged magnetization $\bar{M}_{x}^{(\pm)}$ in terms of derivatives of the quasienergies as follows:
\begin{align}
      \label{AveragedTransvMagnetization}
           \bar{M}_{x}^{(\pm)}&=-\int_{0}^{T}\frac{dt}{T}\int_{0}^{\pi}\frac{dk}{2\pi}\left[\Psi^{(\pm)}_{k,m}(t)\right]^{\dagger}\left(\frac{\partial \mathcal{H}_k}{\partial g_0}\right)\Psi^{(\pm)}_{k,m}(t)
           \nonumber \\ 
           &= -\int_{0}^{\pi}\frac{dk}{2\pi}\frac{\partial \varepsilon_{k,m}^{(\pm)}}{\partial g_0}=\mp\int_{0}^{\pi}\frac{dk}{\pi}\cos(2\phi_{k,m})                    
           .      
\end{align}
Figures ~\ref{Fig4} and ~\ref{Fig5} show the cycle-averaged expectation values of observables. 
Similarly to the undriven case, the system exhibits nonanalyticities in the second derivative of observables---as can be seen in the corresponding insets ---thus resembling a continuous phase transition. Such nonanalyticities arise as a consequence of CDT \cite{CDT1,CDT2}, and therefore, from the gapless quasienergy spectrum. 
  
\section{Conclusions\label{SectionIV}}
We have investigated the nonequilibrium critical behavior in a driven one-dimensional transverse Ising model. We find that the fundamental signature of critical behavior is the existence of a gapless quasienergy spectrum, which is a direct consequence of CDT. The role of coherent destruction of tunneling in nonequilibrium QPT has been explored either theoretically or experimentally in the context of driven superfluidity \cite{Holthaus, Arimondo}. In this paper, we show that CDT induces
a critical behavior which resembles a second-order QPT. In particular, the symmetry which is broken corresponds to a generalized parity in an extended Hilbert space $\mathcal{R}\otimes\mathcal{T}$, where $\mathcal{R}$ is the Hilbert space of square integrable functions and $\mathcal{T}$ is the space of time-periodic functions. 
In this nonadiabatic scenario, the short time dynamics is governed by transient effects that tend to zero in the asymptotic limit as a consequence of destructive interference \cite{Santoro}. The long-time dynamic is governed by the Floquet modes, therefore the nonequilibrium quantum phases correspond to states which are synchronized with the external control.

By means of a Kramers-Wannier self-duality transformation \cite{Peshel,Cirac}, it is possible to map Hamiltonian equation \eqref{drivenIsing} into a dual periodically driven Ising model with time-dependent exchange interaction.
Therefore, the study of the nonequilibrium QPT in the Hamiltonian equation \eqref{drivenIsing} allows one to get a physical picture of the corresponding QPT in the dual model.

A possible experimental implementation of our model could be achieved based on a configuration of superconductor
quantum bits with programmable spin-spin interaction \cite{JohnsonRose}; such a setup allows for a high degree of control of the system parameters. We anticipate that under an adiabatic change of the static local field $g_{0}$ and the 
driving amplitude $g_{1}$ our model could be interesting in the context of quantum annealing, as the effective Hamiltonian equation \eqref{mResonaceIsingXY} corresponds to the $XY$ model. Another experimental setup can be realized by means of cold atoms  \cite{Trotzky, Barreiro, Simon}, and in fully C-labeled sodium butyrate using liquid state nuclear magnetic resonance \cite{Khitrin1,Khitrin2}. 
\begin{acknowledgments}
 The authors gratefully acknowledge discussions with A. B. Bhattacherjee, G. Engelhardt, M. Hayn, V. A. Leyton, C. Nietner, M. Plenio, B. Regler, and M. Vogl, and financial support from the DAAD and DFG Grants BR $1528/7-1$, $1528/8-1$, SFB $910$, GRK $1558$, and SCHA $1646/2-1$.
\end{acknowledgments}
%
\appendix
%
\section{Description of the periodically-driven Ising model for finite size \label{AppendixA}}
In this section we introduce the fundamental tools used in the solution of the Ising model following the methods and the notation of Ref. \cite{Dziarmaga}.

Let us consider the Hamiltonian equation \eqref{drivenIsing} in the case of even number of lattice sites $N$. For convenience, we use the Jordan-Wigner representation of the Pauli matrices
\begin{align}
       \label{JordanWigner} 
	     \sigma^{x}_{j} & =1-2c^{\dagger}_j c_j ,
	      \nonumber \\
	      \sigma^{y}_{j} & = -i(c^{\dagger}_j -c_j)\prod^{j-1}_{l=1}(1-2c^{\dagger}_l c_l) ,
	      \nonumber \\
	      \sigma^{z}_{j} &= (c^{\dagger}_j +c_j)\prod^{j-1}_{l=1}(1-2c^{\dagger}_l c_l)
       .
\end{align}
Under this representation of the angular momentum algebra, the parity operator equation \eqref{IsingParity} acquires  the form 
\begin{equation}
      \label{IsingJWParity}
            \hat{\Pi}=\prod^{N}_{j=1}(1-2c^{\dagger}_j c_j).
\end{equation}
Using this parity operator we are able to define projectors on the subspaces with an even $(+)$ and odd $(-)$ number of fermionic quasiparticles as follows
\begin{equation}
      \label{JWProjectors}
            \hat{\mathcal{P}}_{\pm}=\frac{1}{2}(1 \pm \hat{\Pi}).
\end{equation}
The projectors $\hat{\mathcal{P}}_{\pm}$ satisfy the usual properties of orthogonal projection operators such as $\hat{\mathcal{P}}_{+}+\hat{\mathcal{P}}_{-}=1$, $\hat{\mathcal{P}}_{+}\hat{\mathcal{P}}_{-}=\hat{\mathcal{P}}_{-}\hat{\mathcal{P}}_{-}=0$, and $(\hat{\mathcal{P}}_{\pm})^2=\hat{\mathcal{P}}_{\pm}$. Using these properties and the algebra of fermionic operators it is possible to show that the Hamiltonian equation \eqref{drivenIsing}
can be decomposed as follows
\begin{equation}
      \label{IsingSplitting}
            \hat{H}(t)=\hat{\mathcal{P}}_{+}\hat{H}^{(+)}(t)\hat{\mathcal{P}}_{+}+\hat{\mathcal{P}}_{-}\hat{H}^{(-)}(t)\hat{\mathcal{P}}_{-} ,
\end{equation}
where 
\begin{equation}
      \label{IsingJWRep}
            \hat{H}^{(\pm)}(t)=-g(t)\sum_{i=1}^{N}(1-2c^{\dagger}_i c_i)-J\sum_{i=1}^{N}(c^{\dagger}_i -c_i)(c^{\dagger}_{i+1} +c_{i+1})
      .      
\end{equation}
To perform the splitting we have defined antiperiodic boundary conditions in the even $(+)$ subspace $c_{N+1}=-c_1$ and periodic boundary conditions in the odd $(-)$ subspace $c_{N+1}=c_1$. In this paper we focus on the projection $\hat{H}^{(+)}$ in the even subspace. Translational invariance suggests the use of discrete Fourier transform
\begin{equation}
      \label{DiscreteFourier}
            c_n = \frac{ e^{-i\pi/4}}{\sqrt{N}}\sum_k c_k e^{ikn}
      ,
\end{equation}
which is compatible with the antiperiodic boundary conditions when $k \in \{\pm\frac{\pi}{N},\pm\frac{3\pi}{N},\ldots,\pm \frac{(N-1)\pi}{N}\}$. The discrete Fourier transform maps $\hat{H}^{(+)}$ into Hamiltonian equation \eqref{FourierIsing}.

\section{The quantum phase transition in the anisotropic $XY$ spin chain in a transverse field \label{AppendixB}}
Here we consider the critical behavior in a model described by the Hamiltonian
\begin{eqnarray}
      \label{XYModel}
             \hat{H} &=& -h\sum_{i=1}^{N}\sigma_i^x-\sum_{i=1}^N(J_z\sigma_i^z\sigma_{i+1}^z+J_y\sigma_i^y\sigma_{i+1}^y)
      ,
\end{eqnarray}
which is unitarily equivalent to the Hamiltonian of an anisotropic $XY$ spin chain in a transverse field \cite{Mattis, Barouch, Bunder}.

Similarly to the Ising model, after Jordan-Wigner transformation, and a discrete Fourier transform, the Hamiltonian equation \eqref{XYModel} in the even subspace (the subspace with an even number of fermionic quasiparticles) acquires the form 
\begin{eqnarray}
      \label{FourierXY}
             \hat{H}&=& \sum_{k>0}\left\{2\left[ h-(J_z+J_y)\cos k\right](c_k^{\dagger}c_k+c_{-k}^{\dagger}c_{-k})-2h\right\}  
             \nonumber \\&&
             +\sum_{k>0}2 (J_z-J_y)\sin k(c_k^{\dagger}c_{-k}^{\dagger}+c_{-k} c_{k})
      .
\end{eqnarray}
The diagonalization of this Hamiltonian is completed after a Bogoliubov transformation,
\begin{equation}
      \label{XYModelDiagonal}
           \hat{H}=\sum_{k} E_k\left(\gamma_k^{\dagger}\gamma_k-\frac{1}{2}\right)
      ,
\end{equation}
where 
\begin{equation}
E_k=2\sqrt{\left[h-\left(J_z+J_y\right)\cos k\right]^2+\left[\left(J_z-J_y\right)\sin k\right]^2}.
\end{equation}
The system exhibits an Ising-like QPT along the lines $|h|=J_z +J_y$ and an anisotropic QPT along the line $J_z=J_y$, providing that $|h|<J_z +J_y$. The anisotropic transition is characterized by two ferromagnetic phases, i.e., for $J_z > J_y$ the system is in a ferromagnetically ordered phase along the $z$ direction FM$Z$, while it is the other way around in the FM$Y$ phase. Similarly to Ref. \cite{Mattis}, we consider a reparametrization of the asymmetries
\begin{align}
       \label{AsymmetriesReparametrization} 
	     J_z & =\frac{J}{2}(1+\gamma),
	      \nonumber \\
	     J_y & =\frac{J}{2}(1-\gamma)
       ,
\end{align}
where $\gamma$ is a dimensionless parameter characterizing the degree of anisotropy in the $zy$ plane. Under this reparametrization, the Ising-like critical lines correspond to $|h|=J$, and the anisotropic transition occurs at $\gamma=0$, as long as $|h|<J$.

Interestingly, in the absence of a transverse field, i.e., for $h=0$, the scaled ground-state energy can be written in the thermodynamic limit as
\begin{align}
      \label{ScaledGroundStateEnergy}
            E_G & =-\lim_{N\rightarrow \infty}\frac{1}{N}\sum_k \frac{E_k}{2}
                  \nonumber \\
                & = -J\int_{-\pi}^{\pi}\frac{dk}{2\pi}\sqrt{1-(1-\gamma^2)\sin^2 k} 
                  \nonumber \\
                & = -\frac{2J}{\pi}E[1-\gamma^2]
      ,
\end{align}
where $E[z]$ is the complete elliptic integral of the second kind \cite{Abramowitz}. The scaled ground state energy exhibits a nonanalyticity of the second derivative at the critical line of the anisotropic transition $\gamma=0$ \cite{Mattis}, which is a generic characteristic of a second-order QPT \cite{Sachdev}.
\section{Numerical calculation of the expectation values \label{AppendixC}}
By using the BCS ansatz, Eq. \eqref{BCSAnsatz}, we can solve the Schr\"odinger equation for the Ising model in terms of the solution of the Schr\"odinger equation for an effective two-level system described by the BdG Hamiltonian equation \eqref{TimeDepBogoliubovHamiltonian}, which is parametrized by the quasimomentum $k \in \{\pm\frac{\pi}{N},\pm\frac{3\pi}{N},\ldots, \pm\frac{(N-1)\pi}{N}\}$. 

In the numerical calculation we assume that the system is prepared initially in the unoccupied state $|0_{-k},0_{k}\rangle$, which implies that $\Psi_{k}^{\dagger}(0)=(u^{*}_{k}(0),v^{*}_{k}(0))=(0,1)$. After numerical integration of the dynamical BdG equation \eqref{DynBogoliubovdeGennes}, we find the spinor $\Psi_{k}(t)$. 
To calculate the scaled expectation value of the transverse magnetization $M_x(t)=\frac{1}{N} \left\langle \sum_{i=1}^{N}\sigma^{x}_{i}\right\rangle$ for a given system size $N$, we use the formula
\begin{equation}
      \label{MagnetizationFinite}
            M_x(t) =-\frac{2}{N}\sum_{k>0}\left[\Psi_{k}(t)\right]^{\dagger}\sigma^{z}(k)\Psi_{k}(t)
      .
\end{equation}
In the last expression we have used the definition of $\sigma^{z}(k)$  given in equation \eqref{SigmaZNambu}. For example, to calculate the dynamics of the system for $N=4$, we perform the numerical integration of equation \eqref{DynBogoliubovdeGennes} for $k \in \{\frac{\pi}{4},\frac{3\pi}{4}\}$. Based on the solution of this equation we find the solution of the Schr\"odinger equation for Hamiltonian equation \eqref{drivenIsing} using the BCS ansatz
\begin{equation}
      \label{Example}
            |\psi,t\rangle =|\psi_{\pi/4},t\rangle\otimes|\psi_{3\pi/4},t\rangle 
      ,
\end{equation}
where
\begin{equation}
      \label{Example1}
            |\psi_{\pi/4},t\rangle = u_{\pi/4}(t)|1_{-\pi/4},1_{\pi/4}\rangle+v_{\pi/4}(t)|0_{-\pi/4},0_{\pi/4}\rangle
       ,
\end{equation}
and
\begin{equation}
      \label{Example2}
            |\psi_{3\pi/4},t\rangle = u_{3\pi/4}(t)|1_{-3\pi/4},1_{3\pi/4}\rangle+v_{3\pi/4}(t)|0_{-3\pi/4},0_{3\pi/4}\rangle
       .
\end{equation}

\end{document}